\documentclass{article}

\usepackage{arxiv}

\usepackage[utf8]{inputenc} 
\usepackage[T1]{fontenc}    
\usepackage{hyperref}       
\usepackage{url}            
\usepackage{booktabs}       
\usepackage{amsfonts}       
\usepackage{nicefrac}       
\usepackage{microtype}      
\usepackage{lipsum}		
\usepackage{graphicx}
\usepackage{natbib}
\usepackage{doi}

\title{Quantum Neural Network applications to Protein Binding Affinity Predictions}

\author{ 
\href{https://orcid.org/0000-0003-0520-5075}{\includegraphics[scale=0.06]{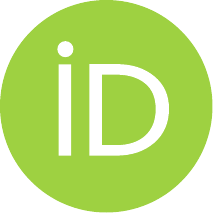}\hspace{1mm}Erico Souza Teixeira} \\
	Excellence Center in Quantum Computing \\
    Venturus\\
    Campinas, SP, Brazil \\
    \texttt{erico.teixeira@venturus.org.br}
\And
\href{https://orcid.org/0009-0005-4690-5600}{\includegraphics[scale=0.06]{orcid.pdf}\hspace{1mm}Lucas Barros Fernandes} \\
Department of Computer Science \\
CESAR School \\
Recife, Brazil \\
\texttt{lbf2@cesar.school}
\And
\href{https://orcid.org/0009-0000-4340-4652}{\includegraphics[scale=0.06]{orcid.pdf}\hspace{1mm}Yara Rodrigues Inácio} \\
Department of Computer Science \\
CESAR School \\
Recife, Brazil \\
\texttt{yri@cesar.school}
}

\date{}

\renewcommand{\shorttitle}{\textit{arXiv}}

\hypersetup{
pdftitle={Quantum Neural Network applications to Protein Binding Affinity Predictions},
pdfsubject={q-bio.NC, q-bio.QM},
pdfauthor={Erico Souza Teixeira, Lucas Barros Fernandes, Yara Rodrigues Inácio},
pdfkeywords={Quantum Neural Networks, Binding affinity, Quantum Computing, Quantum regression},
}

\begin{document}
\maketitle

\begin{abstract}
Binding energy is a fundamental thermodynamic property that governs molecular interactions, playing a crucial role in fields such as healthcare and the natural sciences. It is particularly relevant in drug development, vaccine design, and other biomedical applications. Over the years, various methods have been developed to estimate protein binding energy, ranging from experimental techniques to computational approaches, with machine learning making significant contributions to this field. Although classical computing has demonstrated strong results in constructing predictive models, the variation of quantum computing for machine learning has emerged as a promising alternative. Quantum neural networks (QNNs) have gained traction as a research focus, raising the question of their potential advantages in predicting binding energies. To investigate this potential, this study explored the feasibility of QNNs for this task by proposing thirty variations of multilayer perceptron-based quantum neural networks. These variations span three distinct architectures, each incorporating ten different quantum circuits to configure their quantum layers. The performance of these quantum models was compared with that of a state-of-the-art classical multilayer perceptron-based artificial neural network, evaluating both accuracy and training time. A primary dataset was used for training, while two additional datasets containing entirely unseen samples were employed for testing. Results indicate that the quantum models achieved approximately 20\% higher accuracy on one unseen dataset, although their accuracy was lower on the other datasets. Notably, quantum models exhibited training times several orders of magnitude shorter than their classical counterparts, highlighting their potential for efficient protein binding energy prediction.
\end{abstract}


\section{Introduction}
Protein-protein interactions play a central role in numerous biomedical applications, from drug and vaccine development to biosensor technology. Proteins can act as inhibitors of harmful molecules, such as viruses and bacteria, making them essential targets for designing antiviral and antibacterial therapies \cite{b1, b2}. They are also critical components in biosensors—analytical devices that detect physiological signals, such as blood metabolite concentrations, and convert them into electrical signals for real-time health monitoring \cite{b6}. All these applications depend on the development of biomaterials capable of binding selectively and strongly to specific target proteins \cite{b5}.

A key factor in designing such biomaterials is understanding the strength of interaction between two molecules. This interaction strength is quantified by the binding energy, also referred to as the Gibbs free energy ($\Delta G$) of binding, which reflects the tendency of two compounds to associate or dissociate \cite{b1}. Accurate estimation of binding energy is therefore fundamental to predicting and optimizing protein-protein interactions, ultimately guiding the discovery of new therapeutic agents and diagnostic tools \cite{b3}.

Due to its critical role in protein study and modeling, several methods have been developed for calculating the binding energy $\Delta{G}$. Among these, experimental methods stand out as they assess variations in different thermodynamic properties to determine the binding affinity $k_d$. Examples of such experimental approaches include isothermal titration calorimetry \cite{b18} and differential scanning calorimetry \cite{b19}.

However, studying, screening, testing, and designing biological materials through experimental methods is highly time-consuming and requires substantial financial investment. In drug discovery, for example, it typically takes 10 to 15 years for a new drug to reach pharmaceutical distribution \cite{b5}, with total development costs often exceeding 2 billion dollars \cite{b4}.

Taking this into consideration, techniques to accelerate the discovery of new ligands are essential for reducing costs and development time. Computational methods capable of rapidly estimating binding energies can significantly aid these processes by predicting the binding affinity of compounds to specific target molecules, thus facilitating the identification of promising candidates \cite{b79}.

Since the 1980s, computational approaches for analyzing and identifying molecules have demonstrated promising results, significantly accelerating the development of new biomaterials \cite{b3}. Notable success cases include the discovery of an inhibitor for the severe acute respiratory syndrome (SARS) virus that spread in China in 2003 and developing drugs targeting breast, prostate, and lung cancer cells \cite{b20}.

Among computational methods, Machine Learning (ML)-based approaches have gained significant traction, improving the entire process of predicting binding energy between molecules. These ML models utilize scoring functions, a supervised learning technique that identifies patterns in molecular data using output values derived from experimental binding energy predictions. A variety of scoring functions have been developed for different models, many of which outperform traditional scoring functions in terms of accuracy and efficiency \cite{b3, b8}.

Building on the success of computer-aided methods for binding energy prediction, quantum computing has emerged as a promising approach to overcoming computational bottlenecks. By harnessing the principles of quantum mechanics, it offers the potential for quantum advantage, addressing optimization and data analysis tasks that are intractable for classical methods \cite{b10, b11}.

Within this context, quantum machine learning (QML) integrates quantum computing with classical machine learning to exploit the strengths of both paradigms \cite{b43}. Recent studies have shown that QML models can match or even surpass classical approaches in predictive accuracy, particularly when leveraging rich structural information. For instance, \cite{b13} proposed a hybrid quantum-classical model that integrates 3D convolutional neural networks (3D-CNNs) and spatial graph CNNs (SG-CNNs) to process three-dimensional molecular structures of protein-ligand complexes, achieving a 6\% improvement in binding affinity prediction over a classical model. This demonstrates the potential of QML to exploit high-dimensional structural features more effectively than classical methods, suggesting that it may uncover complex, previously inaccessible relationships within biomedical data and motivating further investigation into its applicability to molecular interaction studies.

With this in mind, the primary objective of this study is to evaluate whether existing quantum neural network (QNN) architectures can outperform state-of-the-art classical models when applied to different types of data representations. Unlike Domingo et al., who leveraged detailed 3D structural information of protein-ligand complexes, our approach relies exclusively on physicochemical properties of molecules.

This choice offers several practical advantages. Physicochemical descriptors are computationally inexpensive to calculate, enabling high-throughput screening of large molecular libraries. They do not depend on high-resolution 3D structures, which are often unavailable in early-stage drug discovery, and they are less sensitive to conformational variability, reducing structural noise. Furthermore, these descriptors are interpretable and have well-established relationships with molecular interactions, which may help QNNs uncover meaningful patterns. By directly comparing QNNs with the best-performing classical neural network, which has already achieved excellent results in classical mode, we aim to assess whether QML can extract additional predictive power, offering a complementary and more scalable alternative to 3D-based approaches for binding energy prediction.

The remainder of this paper is organized as follows: Section \ref{sec:background} provides an overview of the fundamental concepts necessary for understanding the information presented in this study. Section \ref{sec:data} details the datasets employed in this research, along with the data preprocessing techniques applied. Section \ref{sec:method} outlines the methodology adopted to achieve the study’s objectives. Section \ref{sec:experiments} describes the experiments conducted, including the architectures, quantum encodings, and ansatze utilized. Section \ref{sec:results} presents the results and discussion obtained from the experimental analysis. Finally, Section \ref{sec:conclusion} offers concluding remarks.

\section{Background}
\label{sec:background}

Accurate prediction of protein binding energy is a fundamental challenge in drug discovery and molecular design. This section provides the theoretical and computational background relevant to this study. We begin by defining binding energy and its relation to molecular interactions, then review classical machine learning. Next, we introduce the key concepts of quantum computing that enable quantum machine learning (QML), with an emphasis on variational quantum algorithms (VQAs). Finally, we discuss quantum neural networks (QNNs), which extend these principles to deep learning, motivating the approach explored in this work.

\subsection{Binding Energy}
\label{subsec:BE}

Binding ($\Delta G$) energy is the minimum energy required to separate a bonded molecular compound into its constituent molecules. It represents the energy difference between the intact compound and its components \cite{b48}. This thermodynamic property is fundamental in molecular design, ensuring the desired interactions, and is often quantified as binding affinity ($k_D$), which is directly related to binding energy \cite{b3}. In protein-ligand interactions, binding affinity depends on the balance between association and dissociation rates, reflecting the dynamic equilibrium between the protein (P), ligand (L), and their complex (PL) \cite{b1}.

\subsection{Machine Learning}
\label{subsec:ML}

Machine learning (ML) focuses on developing algorithms that identify patterns in data and make predictions or decisions based on statistical models \cite{b22}, and has been categorized into three groups. In supervised learning, the goal is to estimate a function that maps input variables to known outputs, enabling prediction and classification tasks. Unsupervised learning, on the other hand, discovers hidden structures and relationships within data without the need for predefined labels. Reinforcement learning involves agents interacting with an environment to optimize their behavior through trial and error, guided by a reward system \cite{b22, b62}.

The machine learning workflow consists of three key stages: data preprocessing, model training, and performance evaluation. Preprocessing ensures data quality and suitability for modeling by applying techniques such as normalization to scale inputs \cite{b81}. Model training involves optimizing parameters to learn patterns from the data. Performance evaluation assesses model accuracy using metrics such as the root mean square error (RMSE) for regression tasks, which measures the difference between predicted and actual values \cite{b82}. Effective implementation requires careful data handling and iterative model refinement to enhance predictive performance.

\subsubsection{Neural Networks}
\label{subsec:NN}

A neural network consists of perceptrons—processing units inspired by biological neurons \cite{b49}. Each perceptron receives inputs with associated weights, computes a weighted sum, and adds a bias for adjustment. The result is then passed through an activation function, introducing non-linearity to the model and enabling it to capture complex relationships between inputs and outputs \cite{b50}. Neural networks are structured in layers, typically comprising an input layer, one or more intermediate layers (if applicable), and an output layer. When a model includes multiple intermediate layers, it is referred to as a deep neural network \cite{b30}. Training these networks involves backpropagation \cite{b51}, an iterative process that adjusts the weights to minimize errors and enhance the model's predictive performance.

\subsection{Quantum Computing}
\label{subsec:QC}

Quantum computing combines principles from quantum physics with computer science, offering a new paradigm for information processing by leveraging phenomena such as superposition, entanglement, and interference \cite{b31}. These concepts open up the possibility for quantum systems to handle complex problems such as optimization \cite{b32} and machine learning \cite{b34} more efficiently than classical approaches. Unlike classical bits, which represent binary states (0 or 1), qubits can exist in a superposition of these states. This allows quantum systems to process multiple states simultaneously, offering significant performance gains \cite{b35}.

Entanglement is a further cornerstone of quantum computing, where qubits become interconnected so that their states cannot be described independently. This phenomenon enables parallel computation and enhances quantum algorithms. Interference is a distinct key principle in which the probability amplitudes of different quantum states combine, either reinforcing each other (constructive interference) or canceling each other out (destructive interference). These foundational principles not only distinguish quantum computing from classical systems but also drive its revolutionary potential across diverse fields.

\subsubsection{Quantum Circuits}
\label{subsec:QCircuits}

Quantum circuits provide a visual and mathematical framework for modeling computations on qubits. They are composed of sequences of quantum gates that manipulate qubit states through unitary transformations. In contrast to classical logic gates, which operate deterministically on binary values, quantum gates exploit phenomena like superposition and interference, enabling complex state evolution and unlocking computational capabilities unique to quantum systems.

Multi-qubit gates, such as the controlled-NOT (CNOT), facilitate the generation of entanglement, forming the backbone of quantum operations. Rotation gates allow for precise manipulation of individual qubit states on the Bloch sphere, enabling arbitrary single-qubit transformations. Controlled rotation gates combine the principles of entanglement and conditional logic, enabling more expressive and flexible quantum circuit designs essential for quantum algorithms.

Parameterized Quantum Circuits (PQCs) extend this framework by introducing tunable parameters that can be optimized for specific tasks. Rather than relying on fixed operations, PQCs use parameter-dependent gates to dynamically explore a broader range of quantum states.

These circuits commonly incorporate gates like $R_x(\theta)$, $R_y(\theta)$, and $R_z(\theta)$, which perform rotations around the respective axes of the Bloch sphere, guided by input parameters $\theta$. Controlled variants (e.g., $CR_x$, $CR_y$, and $CR_z$) apply these rotations conditionally, depending on the state of another qubit, thereby enabling entanglement and supporting more intricate quantum behavior.

\subsubsection{Measurement}
\label{subsec:MM}
Measurement is the final and crucial step in quantum circuits, where quantum information is projected onto classical states. In this process, qubits collapse from a superposition of basis states into definite classical values, either 0 or 1, with probabilities determined by the squared magnitudes of their respective amplitudes. This probabilistic nature of quantum measurement reflects the fundamental principles of quantum mechanics, where only statistical properties of the system can be inferred through repeated observations.

Since a single measurement provides only one possible outcome based on the inherent quantum uncertainty, multiple runs of the same circuit are necessary to build a probabilistic distribution of results. This repeated sampling allows for accurate estimations of quantum state properties and expectation values of observables \cite{b54}.

\subsection{Variational Quantum Algorithms}
\label{subsec:VQA}

Variational Quantum Algorithms (VQAs) are among the leading approaches for quantum computing on noisy quantum devices. They mitigate noise errors by leveraging a hybrid framework, where classical computers optimize parameters while quantum circuits with shallow depth execute computations \cite{b37}\cite{b86}\cite{b87}. Based on the variational principle, which seeks to determine the lowest energy state of a system, VQAs iteratively adjust the parameters of a quantum wave function to minimize the system's energy, using a Hamiltonian operator \cite{b56}.

A VQA comprises three main components: a Parameterized Quantum Circuit (PQC), or ansatz, which represents the quantum wave function; a cost function, or Hamiltonian, which encodes the problem as a measurable quantity; and a classical optimizer, which updates the ansatz parameters to minimize the cost function \cite{b37}.

\subsection{Quantum Machine Learning}
\label{subsec:QML}

Quantum Machine Learning (QML) is an emerging interdisciplinary field that integrates quantum computing principles with classical machine learning techniques to exploit the computational advantages of quantum hardware. By leveraging quantum superposition, entanglement, and interference, QML aims to enhance machine learning tasks, particularly in scenarios where classical methods struggle with scalability and efficiency, such as training large-scale models and high-dimensional data processing \cite{b31}\cite{b43}. This field encompasses both quantum-enhanced adaptations of traditional machine learning models, such as quantum versions of support vector machines, convolutional neural networks, and kernel methods \cite{b46}\cite{b47}, and the broader exploration of how data can be represented and processed within quantum systems \cite{b43}.

This approach can bring benefits when compared to classical methods, such as:
\begin{enumerate}
    \item QML models have the potential to identify patterns that classical models may overlook, leading to more accurate predictions.
    \item Quantum computers, with their higher capacity for information representation and reduced computational resource requirements, can leverage parallel processing enabled by quantum superposition.
\end{enumerate}

\subsubsection{Feature Encoding}
\label{subsec:DE}
A fundamental aspect of QML is feature encoding, which involves transforming classical or quantum data into quantum states that can be efficiently manipulated by quantum algorithms. The choice of encoding strategy has a significant impact on the expressiveness and efficiency of quantum models. Common encoding techniques include:

\begin{enumerate}
    \item Angle Encoding: Maps classical data features onto qubit states using rotational transformations.
    \item Amplitude Encoding: Normalizes classical data and embeds it into the probability amplitudes of quantum states, enabling efficient representation of high-dimensional vectors within exponentially large Hilbert spaces \cite{b44}.
    \item Basis Encoding: Directly associates classical bitstrings with computational basis states, commonly used in binary classification tasks.
    \item Hybrid Encoding: Combines multiple encoding strategies to optimize expressiveness and resource efficiency.
\end{enumerate}

\subsubsection{Quantum Neural Networks}
\label{subsubsec:QNN}

Quantum Neural Networks (QNNs) extend variational quantum algorithms to deep learning by leveraging parameterized quantum circuits for tasks such as energy minimization, optimization, and pattern recognition \cite{b61}. Unlike classical neural networks, which rely on matrix multiplications and activation functions, QNNs encode input data into quantum states through feature encoding, manipulate these states using quantum gates, and optimize the circuit parameters—analogous to weights and biases—through classical optimization techniques like gradient-based methods and variational parameter tuning \cite{b11}.

Hybrid Quantum Neural Networks (HQNNs) combine quantum and classical computational layers, leveraging the strengths of both paradigms to enhance deep learning applications. In HQNN architectures, classical data is first mapped into quantum states via feature embedding, processed using parameterized quantum circuits, and finally measured to extract outputs fed into classical layers for further analysis and refinement \cite{b68}. This hybrid approach mitigates current quantum hardware limitations, such as noise and decoherence, while still benefiting from quantum-enhanced 
computation in key subroutines.

HQNNs have shown promise in applications requiring high-dimensional feature representations, such as optimization, anomaly detection, and signal processing \cite{b68}. By integrating classical deep learning with quantum resources, HQNNs facilitate efficient quantum data processing while maintaining compatibility with existing machine learning frameworks \cite{b68}. 

\section{Data}
\label{sec:data}

Due to the comparative nature of this study, which evaluates classical machine learning models against quantum machine learning (\textit{QML}) approaches, the selected datasets for the experiment were the same as those used in the study by Ferraz \textit{et al.} \cite{b3}. One dataset was chosen for model training, while two additional datasets were used to assess the generalization capability of the proposed models.

To ensure the highest level of similarity between the input data used in the classical model and the data introduced in the present study, the same data preprocessing methodology was applied. It is important to emphasize that these procedures were performed on the classical data before transforming it into the quantum context.

The preprocessing steps were as follows: first, the dataset was cleaned by removing anomalous samples; next, the values were normalized to a range of 0 to 1; and finally, we applied Uniform Manifold Approximation and Projection (UMAP) for dimensionality reduction, decreasing the original 41 features to 16. This dimensionality was selected because it achieved the best predictive performance for the baseline classical model while reducing overfitting.

The first dataset, used for training, was initially extracted from the research conducted by Vangone and Bonvin \cite{b70}. It consists of 81 protein structure samples with known binding energy values. Despite its relatively small size, this dataset encompasses a diverse range of protein interfaces. The exploratory data analysis of this training dataset is detailed in Ferraz \textit{et al.} \cite{b3}, where correlations between sample features and binding energy values are examined.

One of the datasets selected for model validation comprises 38 crystal structure complexes of nanobodies (also known as \textit{nbs}) with known binding energy values. This dataset was curated from a more extensive collection of 123 non-redundant structures available in \textit{pdb} format \cite{b73}.

The second dataset used for model validation consists of 50 protein complexes with known binding energy values, selected from the \textit{PDBind} database—an online-accessible repository containing 1,359 protein complexes \cite{b72}. The selection and filtering process applied to the \textit{PDBind} dataset to derive the 50 samples used in this study is described in Ferraz \textit{et al.} \cite{b3}.

\section{Materials and Methods}
\label{sec:method}

\begin{figure}[h]
    \centering
    \includegraphics[width=1\linewidth]{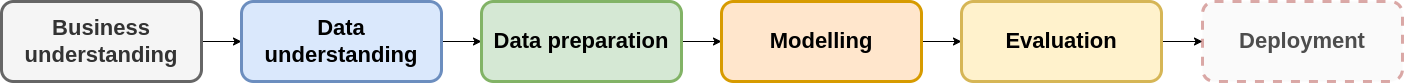}
    \caption{Steps of the methodology used}
    \label{fig:method}
\end{figure}

The research methodology was adapted from the Cross-Industry Standard Process for Data Mining (CRISP-DM), a widely used framework for machine learning projects \cite{b67}. Initially proposed by Wirth and Hipp (2000), CRISP-DM 
outlines the lifecycle of data-driven model development. As illustrated in Figure \ref{fig:method}, the deployment phase is excluded, indicated by shading and dashed borders, since this study is conducted in an academic setting. Model deployment involves implementing the trained model for an end-user application, which is beyond the scope of this research.

This research follows a structured approach comprising five key phases: business understanding, data understanding, data preparation, model application, and result evaluation. The business understanding phase establishes the study’s objectives by analyzing the domain where machine learning can be applied, specifically focusing on contextualizing the binding energy prediction task \cite{b67}. 

The data understanding phase encompasses data collection, identification of data types, and exploratory analysis to reveal potential correlations within the dataset \cite{b67}. However, since this study builds upon the work of 
Ferraz, Neto, Lins, and Teixeira \cite{b3}, which includes a comprehensive exploratory analysis, did not replicate this phase in full. Instead, the existing insights from their study were leveraged to inform subsequent steps.

Data preparation is a critical step, especially when the input data for QML models is classical. Adapting the data for quantum use requires careful preprocessing before transforming it into quantum states. The preprocessing steps followed in this research are better explained in Section \ref{sec:data}. It is essential to note that the study conducted by Ferraz et al. included a dimensionality reduction step, in which the number of features in the datasets was reduced from 41 to 15. However, in the present work, the number of features was reduced from 41 to 16, as this dimensionality was considered more suitable for performing the feature encoding process.

As already mentioned in Section \ref{sec:background}, feature encoding is the process of converting classical values into quantum states so that quantum systems can manipulate them. There are various feature encodings available in the literature \cite{b44}; however, the focus of this study is restricted to two types: angle encoding and amplitude encoding, with both methods applied after data cleaning, normalization, and dimensionality reduction, ensuring compatibility with the constraints of quantum information processing.

Angle encoding involves converting classical information into rotation angles of quantum gates around an axis of the Bloch sphere. Since angle encoding uses $n$ qubits to encode $n$ features into quantum states, the number of required qubits can become significantly large depending on the number of features in the dataset. To address this, the data re-uploading technique, developed by Pérez et al. \cite{b76} and adapted by Avramouli et al. \cite{b65}, was utilized.

Using the selected feature set in the experimental study as an example, data re-upload maps the 16 features of each sample into four blocks of four qubits. The first four features are mapped onto the first block, transforming a series of quantum gates that form a parameterized quantum circuit incorporating a set of weights $w$. Afterward, the following four features are applied to the second block via feature mapping, and this process continues until all features have been converted into rotation angles and applied to the qubits.

Amplitude encoding, on the other hand, transforms a set of classical data into probability amplitudes of quantum states in superposition. First, it is necessary to ensure that the feature values are normalized, meaning the sum of the square modulus of these values must equal 1. This is achieved by calculating the norm of these values and then dividing each feature value by the computed norm \cite{b44}. This type of encoding requires only $log_2(n)$ qubits to convert $n$ features into quantum states, as it embeds the values of the independent variables into the probability amplitudes of an arbitrary quantum state $|\psi\rangle$.

The encoding strategies employed were deliberately selected because they are widely used in QML literature and compatible with near-term hardware constraints. Angle encoding with re-uploading enables multiple feature insertions with shallow circuits, which is advantageous for NISQ devices, even though its expressivity is limited when using few re-uploads. Amplitude encoding was included to assess the performance of a compact embedding with exponentially reduced qubit requirements, despite its mostly linear representational capacity when followed by shallow PQCs. These choices were intended to establish a reproducible baseline before investigating task-specific encodings optimized for expressivity and hardware efficiency.

The final phase, evaluating results, involves assessing the performance of the quantum models using established metrics and conducting a comparative analysis with classical neural networks. The QNN architectures explored in this study are inspired by recent advances in the literature on quantum machine learning for molecular and biochemical systems, particularly the works of Banerjee et al. \cite{b13} and Avramouli et al. \cite{b65}, which demonstrate the applicability of hybrid quantum models in binding energy prediction tasks. These studies provided foundational insights into encoding schemes, circuit design, and integration of quantum layers into learning pipelines, guiding the development of the models evaluated in this research.

For the performance analysis of the quantum models and their comparison with the classical model, the Root Mean Square Error (RMSE) was employed. It is important to highlight that this metric was the sole criterion chosen for evaluating predictive accuracy in the present study, as Ferraz et al. \cite{b3} also employ this metric to report the performance of the classical neural network. RMSE computes the square of the difference between the actual values from the dataset samples and the values predicted by the model. These squared differences are then summed, and the square root of the total is calculated. As this metric quantifies the discrepancy between the predicted and true values, lower RMSE values are desirable, indicating higher predictive accuracy of the model.

All experiments were conducted in a cloud-based environment provided by the Google Colab platform, which offers access to computational resources such as CPUs and GPUs for remote code execution via a web browser. The training procedures were entirely carried out using an NVIDIA L4 GPU, equipped with 16 gigabytes of memory. The decision to run all models on this virtual machine aimed to ensure a consistent and controlled environment for analyzing execution time performance, maintaining identical hardware specifications throughout the process.

It is also important to note that all experiments were executed using a quantum simulator named default.qubit, provided by the PennyLane framework (version 0.38.0), meaning that no computations were performed on actual quantum hardware. This choice was made to focus on evaluating the models’ representational capacity without the confounding effects of hardware noise. However, it should be noted that this setup results in training times that are not representative of realistic NISQ hardware performance. Additionally, PyTorch (version 2.5.1), a well-established machine learning library, was utilized to interface with PennyLane to execute quantum circuits.

The QNN architectures, ansätze, and encoding schemes used in this study were intentionally selected from well-established QML literature rather than designed specifically for this task. This choice was motivated by the exploratory nature of this work, which seeks to provide a fair and interpretable baseline comparison between standard QNN templates and a strong classical neural network benchmark. By starting with these widely adopted circuits and encodings, we aim to assess whether current QML practices can already yield competitive results before exploring task-specific circuit optimizations in future work.

\section{Experiments}
\label{sec:experiments}
The following section introduces the different variations of quantum neural networks employed in the experimental study using the selected datasets. Additionally, the ansätze utilized to represent the hidden layers of the quantum neural networks are outlined.

\subsection{Model architectures}

\begin{figure}[htbp]
\centerline{\includegraphics[width=0.3\textwidth]{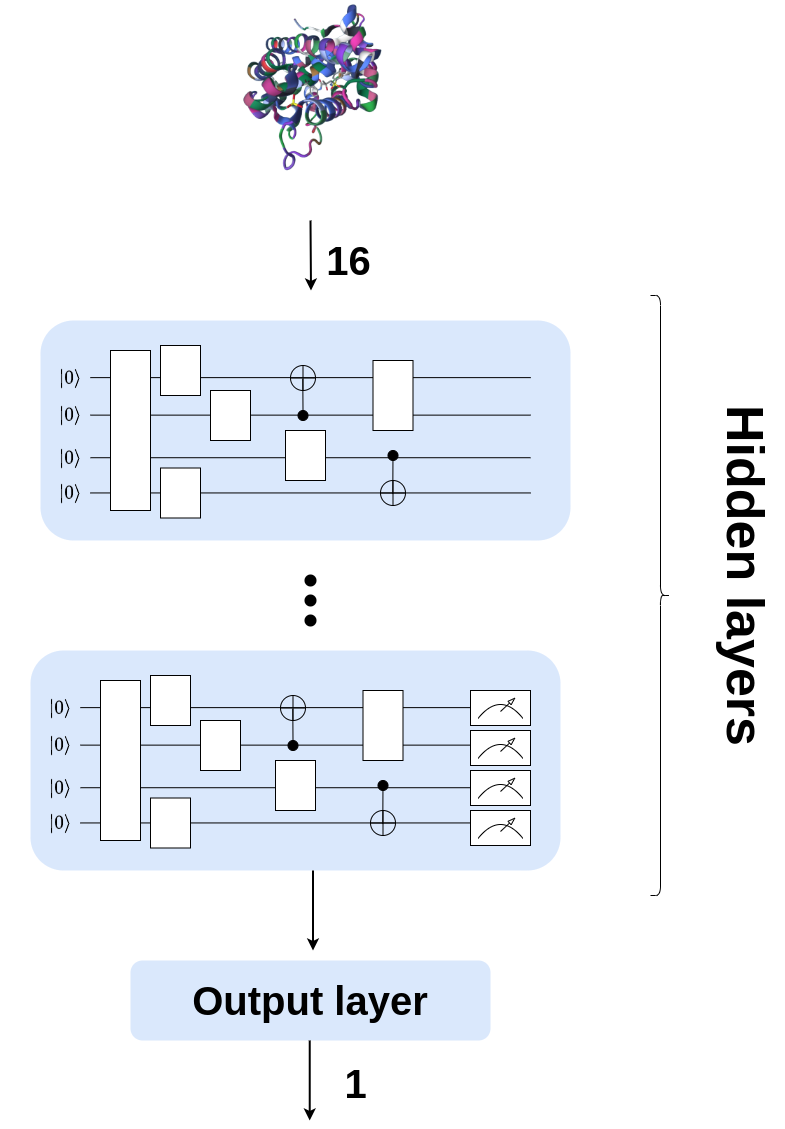}}
\caption{Proposed sequential architecture}
\label{fig:qnn-sequential}
\end{figure}

Three distinct architectures for quantum neural network models were employed, each utilizing five ansätze and two distinct feature encodings, resulting in 30 model variations. The first approach is the \textbf{sequential model} that represents the simplest hybrid quantum-classical approach and serves as a baseline architecture. As shown in Figure \ref{fig:qnn-sequential}, the 16 features of the dataset samples are loaded into the model, which performs the appropriate feature encoding to convert classical values into quantum states (angle encoding with data re-uploading or amplitude encoding). After executing all quantum layers (in this particular study, two quantum layers were employed), the expected values of the qubits along the Z-axis are measured and mapped back to classical values. These classical values are then used in a weighted sum with the weights of a final output layer, resulting in the prediction of the binding energy for the given sample. This approach is similar to methods used in binding energy prediction studies by Banerjee et al. \cite{b13} and Avramouli et al. \cite{b65}.

\begin{figure}[htbp]
\centerline{\includegraphics[width=0.5\textwidth]{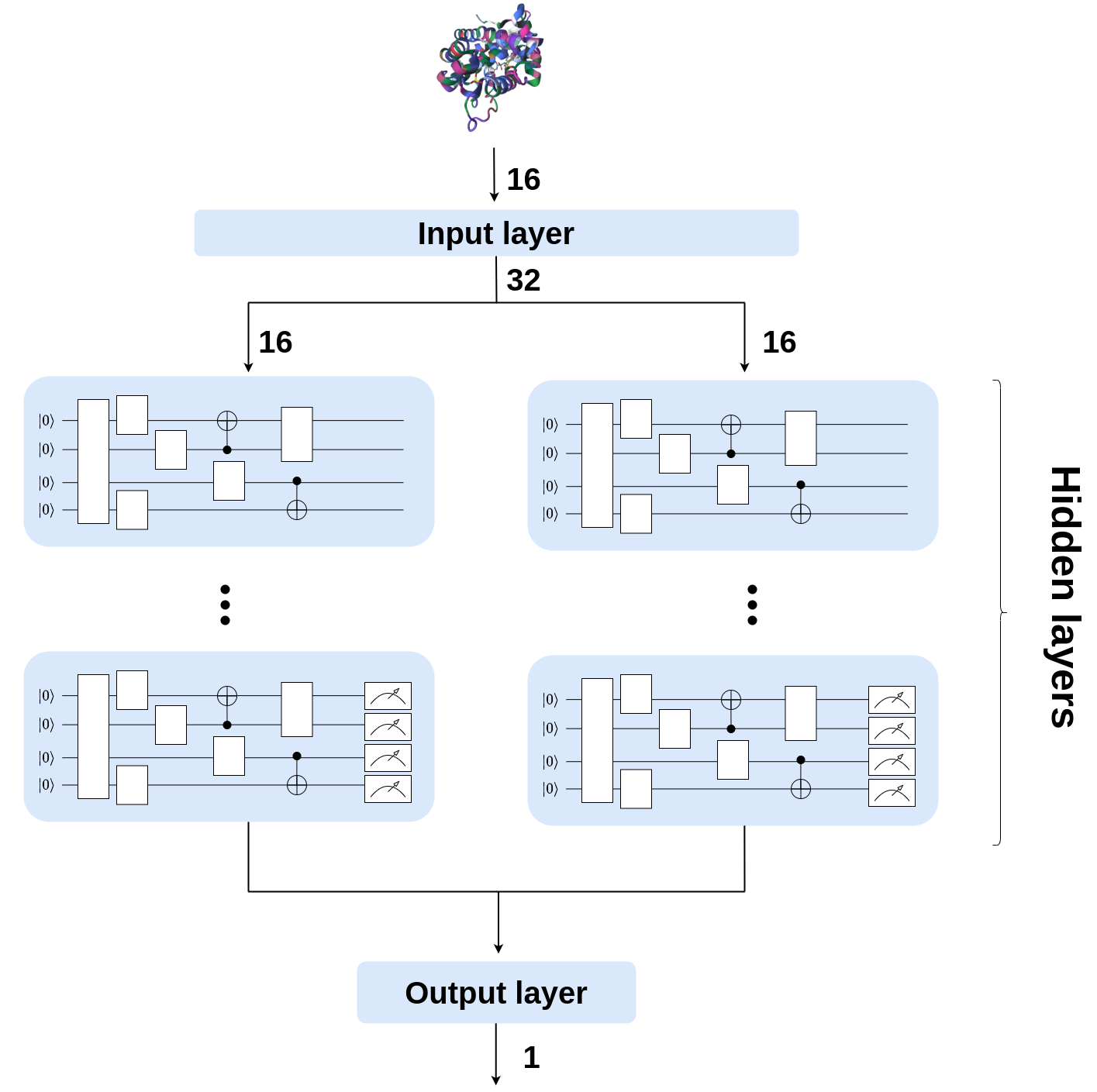}}
\caption{Proposed parallel architecture}
\label{fig:qnn-parallel}
\end{figure}

The second architecture employed is the \textbf{parallel model}, designed to assess whether processing features through independent quantum layers enhances expressivity. Inspired by architectures that process two hidden layers independently and simultaneously, this model allocates 16 qubits to each parallel layer. As shown in Figure~\ref{fig:qnn-parallel}, the 16 input features are first passed through a classical linear layer that expands them to 32 dimensions. Let $\overrightarrow{x} = [x_1, x_2, ..., x_{16}]$ denote a sample's input vector. A weight matrix $\mathbf{W} \in \mathbb{R}^{32 \times 16}$ and a bias vector $\overrightarrow{b} \in \mathbb{R}^{32}$ are used to compute the transformed vector via $\mathbf{y} = \mathbf{x} \cdot \mathbf{W}^T + \mathbf{b}$. The resulting 32-dimensional output $\mathbf{y}$ is then evenly split between the two parallel quantum layers.

Each subset of 16 features is encoded into quantum states using a quantum encoding scheme. The two quantum circuits process their respective quantum states in parallel. After execution, measurement in the computational basis (Z-axis) yields 32 classical outputs, one per qubit. To obtain the final prediction, these values are aggregated through a weighted sum in the output layer, producing a single scalar output.

The third and final architecture adopted is the \textbf{ensemble model} to test whether aggregating predictions from multiple small QNNs can improve generalization, particularly given the small size of the datasets. It consists of a set of 9 sequential models, each trained on nine subsets of the training dataset created through sampling without replacement to generate multiple smaller subsets of the dataset, which are used to train each of the smaller models.

\begin{figure}[!htbp]
\centerline{\includegraphics[width=0.5\textwidth]{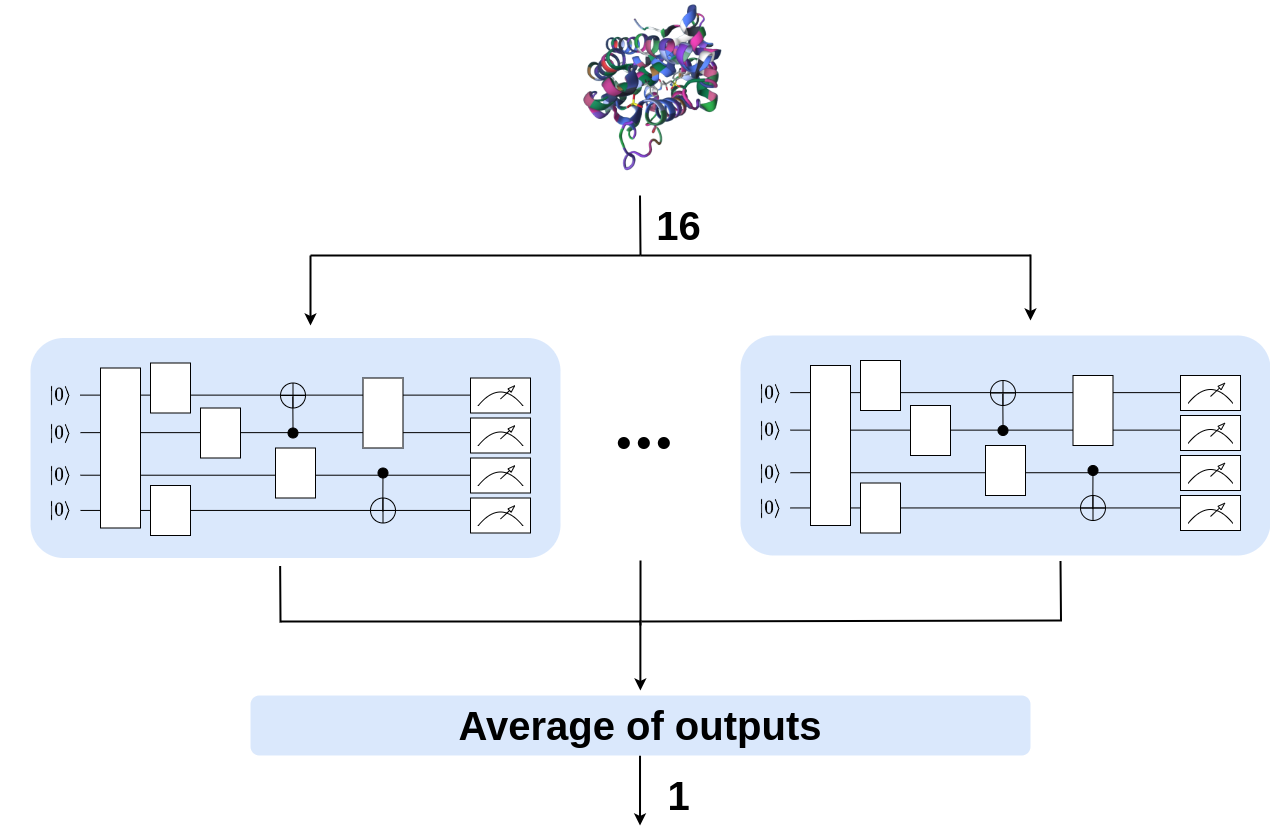}}
\caption{Proposed ensemble architecture}
\label{fig:qnn-ensemble}
\end{figure}

Figure \ref{fig:qnn-ensemble} presents a diagram of this model. For prediction, the 16 features of a sample are loaded into all smaller models, each of which follows the same feature encoding (in this case, angle or amplitude encoding) and parameterized quantum circuit process. These circuits transform the system's state, after which the expected values of the qubits along the Z-axis are measured and subjected to a weighted sum in an output layer, yielding a classical value. Finally, an average is taken over the outputs of the nine smaller models, producing the final prediction.

These three architectures were selected to explore different trade-offs between simplicity, expressivity, and generalization. The sequential model serves as a baseline; the parallel model increases parameterization and expressivity; and the ensemble model mitigates overfitting by combining multiple smaller QNNs, an approach inspired by ensemble methods in classical machine learning.

To ensure fair comparison and stable convergence across all experiments, the QNNs were trained under a standardized optimization protocol. Specifically, the Gradient Descent Estimator (GDE) optimizer was adopted, with a learning rate initialized at 0.01. To prevent overshooting and facilitate convergence toward a local minimum of the cost function, a learning rate scheduler was also employed. This scheduler progressively reduced the learning rate after a defined number of epochs without improvement in model accuracy, allowing finer parameter adjustments during later training stages.

\subsection{Parameterized quantum circuits}

\begin{figure*}[!htbp]
\centerline{\includegraphics[width=1\textwidth]{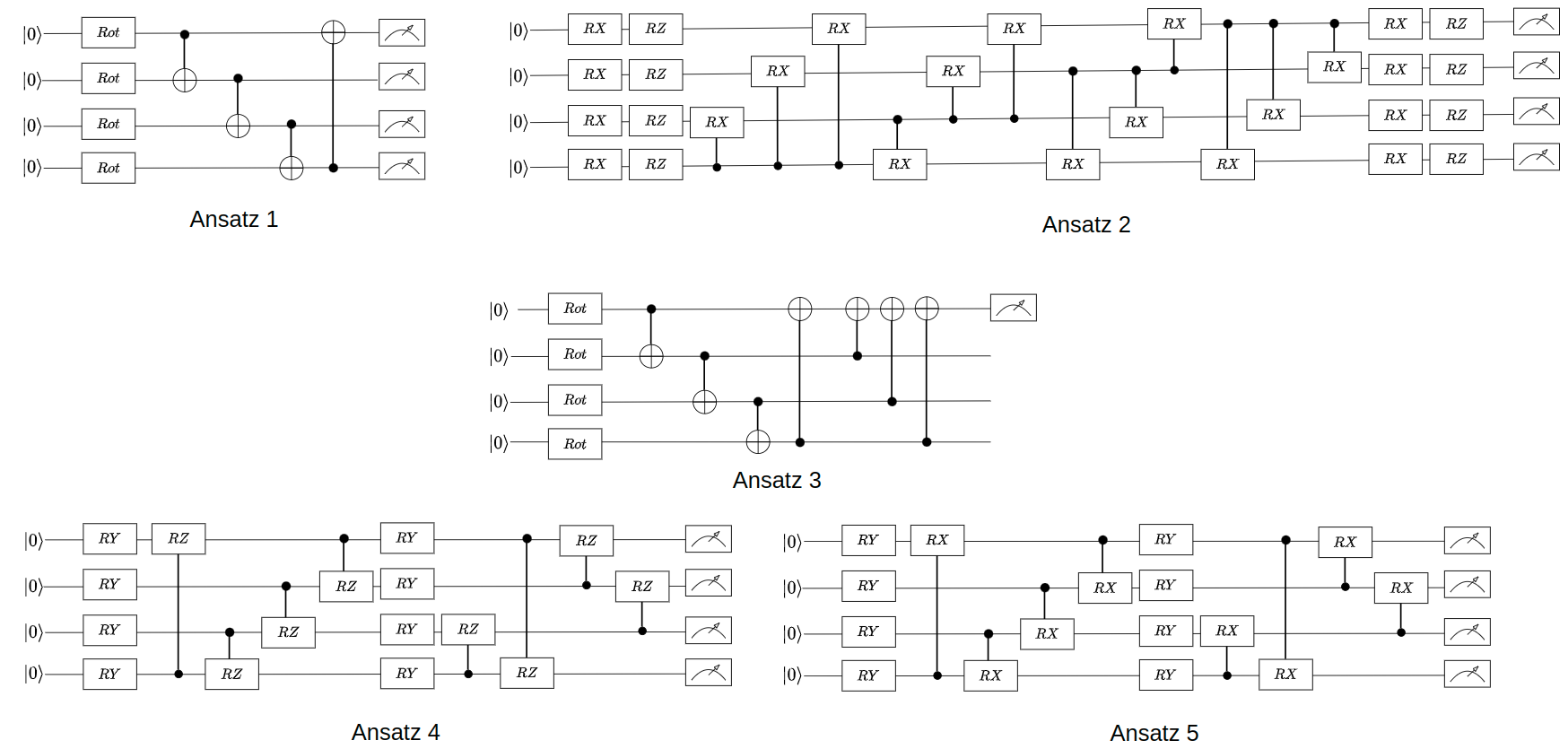}}
\caption{Circuits used on the QNN models}
\label{fig:ansatze}
\end{figure*}

We propose five ansätze that serve as hidden layers of the quantum neural network. We selected these circuits based on their relevance to both the regression problem and protein binding energy prediction work \cite{b13} \cite{b66} \cite{b65}. These parameterized circuits process the quantum states resulting from the feature encoding, utilizing a set of weights to compute the values that will be measured and eventually transformed into the predicted output variable for a given sample. Figure \ref{fig:ansatze} presents the diagrams of the circuits used.

Table \ref{tbl:qnn-ansatz-complexity} presents values related to the complexity of the ansätze used, such as circuit depth, number of 2-qubit gates present, as well as the total number of quantum gates and trainable parameters in each circuit. As already mentioned, two quantum layers were employed for the neural networks.

Due to the use of the data re-upload technique for circuits with angle encoding, it can be observed that the number of quantum gates and the total depth of these circuits are significantly higher compared to those utilizing amplitude encoding. This outcome is expected, as data re-upload processes the neural network's input data across multiple blocks, thereby increasing the number of gates required for its operation.

\begin{table}[ht]
\caption{Ansatz complexity metrics}
\label{tbl:qnn-ansatz-complexity}
\centering
\scriptsize
\begin{tabular}[t]{lccccc}
\hline
\textit{Ansatz}&Layers&Depth&2-qubit gates&\# of gates&Params\\
\hline
Ansatz 1 angle & 2 & 72 & 32 & 72 & 24\\
Ansatz 1 amplitude & 2 & 15 & 8 & 17 & 24\\
Ansatz 2 angle & 2 & 128 & 96 & 232 & 56\\
Ansatz 2 amplitude & 2 & 30 & 24 & 57 & 56\\
Ansatz 3 angle & 2 & 72 & 56 & 96 & 24\\
Ansatz 3 amplitude & 2 & 17 & 14 & 23 & 24\\
Ansatz 4 angle & 2 & 80 & 64 & 136 & 32\\
Ansatz 4 amplitude & 2 & 18 & 16 & 33 & 32\\
Ansatz 5 angle & 2 & 80 & 64 & 136 & 32\\
Ansatz 5 amplitude & 2 & 18 & 16 & 33 & 32\\
\hline
\end{tabular}
\end{table}

The ansätze presented in Figure \ref{fig:ansatze} exhibit distinct trade-offs in terms of expressivity, hardware efficiency, and potential for overfitting. Ansatz 1 is the shallowest, with the lowest gate count and number of parameters, making it more hardware-friendly and less prone to overfitting on small datasets. However, its limited depth may reduce its ability to represent highly nonlinear relationships. Ansatz 2, by contrast, is the deepest and most parameterized, providing greater expressive power but at the cost of higher computational complexity and potentially slower convergence. Ansätze 3, 4, and 5 fall between these two extremes, balancing expressivity and trainability. Interestingly, Ansätze 4 and 5 share similar depth and parameterization, differing mainly in gate arrangement, which may affect entanglement patterns and thus feature interaction modeling.

\section{Results and Discussion}
\label{sec:results}

In this chapter, the results obtained from the models and their previously discussed variations will be presented, emphasizing the impact of different ansätze and feature maps. As mentioned, three different architectures for QNN models were employed, each utilizing five ansätze and two distinct feature maps, resulting in 30 model variations. Tables \ref{tbl:qnn-sequential-comparison}, \ref{tbl:qnn-parallel-comparison}, and \ref{tbl:qnn-ensemble-comparison} present the results of these three modeling approaches, where each row corresponds to a unique combination of ansatz and feature encoding, comprising 10 variations for each model type.

RMSE metrics can also be observed for both the training and testing segments of the training dataset, as well as for the two external datasets, completely unknown to the models, namely Nanobodies (Nbs) and PDBind. The number of parameters for each model variation is also reported. It is important to emphasize that, since the model encompasses both classical and quantum layers, the parameter values presented in the table represent the total number of network parameters, which may belong to either the classical or quantum components.

\begin{table*}[!ht]
\caption{Performance of the sequential model with different ansätze applied to the quantum layer}
\label{tbl:qnn-sequential-comparison}
\centering
\scriptsize
\begin{tabular}[t]{lccccc}
\hline
\textit{Ansatz}&Train - RMSE&Test - RMSE&Nbs - RMSE&PDBind - RMSE&Parameters\\
\hline
Ansatz 1 w/ angle & 4.09 & 3.42 & 4.12 & 2.99&29\\
Ansatz 1 w/ amplitude & 2.99 & 2.73 & 2.60 & 2.04&29\\
Ansatz 2 w/ angle & 5.74 & 5.26 & 5.29 & 4.64&61\\
Ansatz 2 w/ amplitude & 2.96 & 2.70 & 2.54 & 2.00&61\\
Ansatz 3 w/ angle & 4.11 & 4.23 & 4.59 & 3.35&26\\
Ansatz 3 w/ amplitude & 2.92 & 2.64 & 2.14 & 1.92&26\\
Ansatz 4 w/ angle & 6.96 & 6.56 & 7.02 & 5.84&37\\
Ansatz 4 w/ amplitude & 2.89 & 2.61 & 2.30 & 1.92&37\\
Ansatz 5 w/ angle & 5.49 & 4.65 & 5.53 & 5.10&37\\
Ansatz 5 w/ amplitude & 2.92 & 2.67 & 2.47 & 1.97&37\\
\hline
\end{tabular}
\end{table*}

\begin{table*}[!ht]
\caption{Performance of the parallel model with different ansätze applied to the quantum layer}
\label{tbl:qnn-parallel-comparison}
\centering
\scriptsize
\begin{tabular}[t]{lccccc}
\hline
\textit{Ansatz}&Train - RMSE&Test - RMSE&Nbs - RMSE&PDBind - RMSE& Parameters\\
\hline
Ansatz 1 w/ angle & 2.94 & 2.62 & 2.38 & 1.97& 601\\
Ansatz 1 w/ amplitude & 3.00 & 2.77 & 2.66 & 2.06 & 601\\
Ansatz 2 w/ angle & 2.91 & 2.64 & 2.29 & 1.92& 665\\
Ansatz 2 w/ amplitude & 2.95 & 2.71 & 2.40 & 1.96& 665\\
Ansatz 3 w/ angle & 2.90 & 2.64 & 2.23 & 1.93& 595\\
Ansatz 3 w/ amplitude & 3.05 & 2.84 & 2.70 & 2.09& 595\\
Ansatz 4 w/ angle & 2.94 & 2.51 & 2.32 & 1.97& 617\\
Ansatz 4 w/ amplitude & 3.02 & 2.80 & 2.61 & 2.05& 617\\
Ansatz 5 w/ angle & 2.92 & 2.61 & 2.30 & 1.95& 617\\
Ansatz 5 w/ amplitude & 3.05 & 2.82 & 2.84 & 2.15& 617\\
\hline
\end{tabular}
\end{table*}

\begin{table*}[!ht]
\caption{Performance of the ensemble model with different ansätze applied to the quantum layer}
\label{tbl:qnn-ensemble-comparison}
\centering
\scriptsize
\begin{tabular}[t]{lccccc}
\hline
\textit{Ansatz}&Train - RMSE&Test - RMSE&Nbs - RMSE&PDBind - RMSE& Parameters\\
\hline
Ansatz 1 w/ angle & 4.52 & 4.39 & 5.58 & 4.23& 261\\
Ansatz 1 w/ amplitude & 2.98 & 2.74 & 2.56 & 2.01 & 261\\
Ansatz 2 w/ angle & 5.54 & 5.54 & 6.12 & 4.99& 549\\
Ansatz 2 w/ amplitude & 3.01 & 2.79 & 2.66 & 2.06& 549\\
Ansatz 3 w/ angle & 3.79 & 3.67 & 4.07 & 3.17& 234\\
Ansatz 3 w/ amplitude & 2.97 & 2.73 & 2.56 & 2.01& 234\\
Ansatz 4 w/ angle & 4.55 & 4.55 & 4.87 & 3.96& 333\\
Ansatz 4 w/ amplitude & 2.92 & 2.65 & 2.37 & 1.94& 333\\
Ansatz 5 w/ angle & 5.40 & 5.34 & 6.41 & 5.23& 333\\
Ansatz 5 w/ amplitude & 2.94 & 2.67 & 2.47 & 1.98& 333\\
\hline
\end{tabular}
\end{table*}

\begin{figure*}[!htbp]
\centerline{\includegraphics[width=0.9\textwidth]{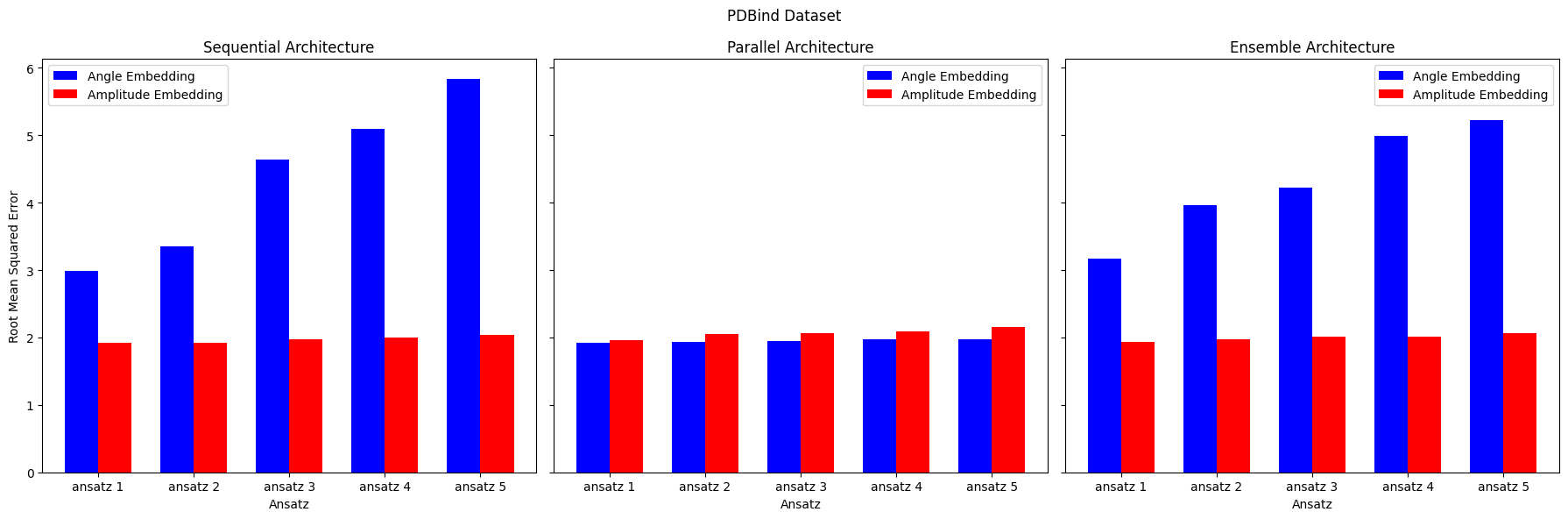}}
\caption{Comparative chart for different ansätze and feature mappings on the PDBind dataset}
\label{fig:pdbind-ansatz-feature-map}
\end{figure*}

\begin{figure*}[!htbp]
\centerline{\includegraphics[width=0.9\textwidth]{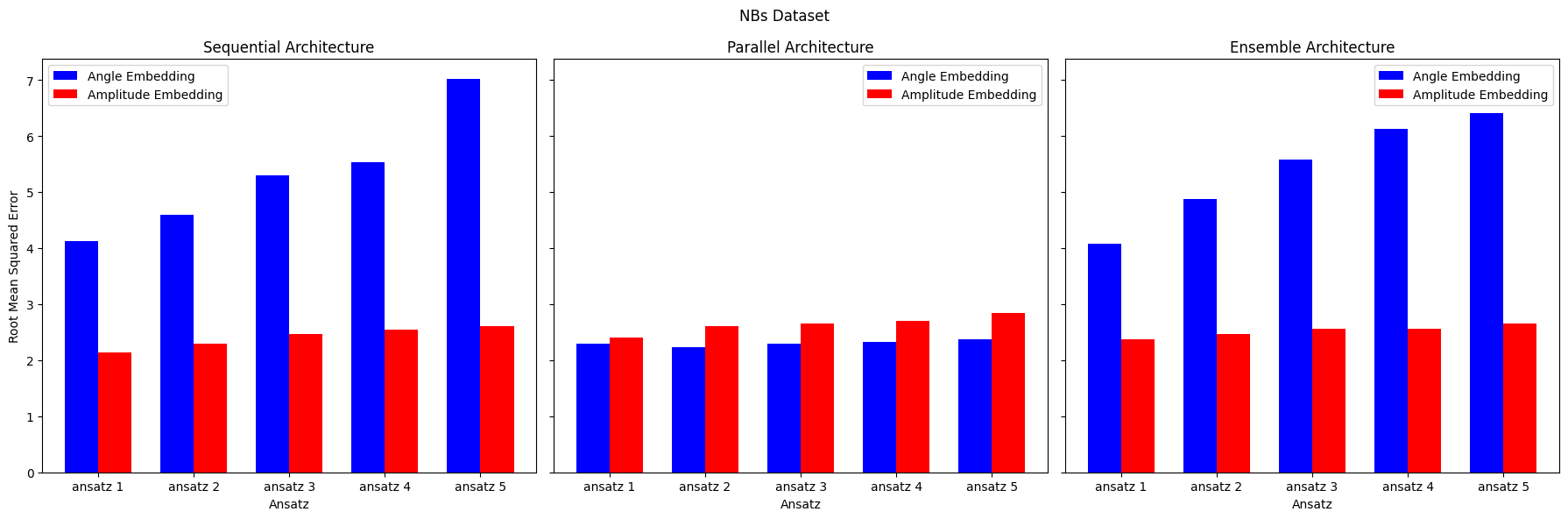}}
\caption{Comparative chart for different ansätze and feature mappings on the Nanobodies dataset}
\label{fig:nbs-ansatz-feature-map}
\end{figure*}

Graph visualizations were conducted to analyze the impact of the different ansätze and feature encodings applied. Figures \ref{fig:pdbind-ansatz-feature-map} and \ref{fig:nbs-ansatz-feature-map} illustrate the RMSE results for different ansatz and encoding pairs for the architectures used in the PDBind and Nanobodies datasets. A significant difference is observed between the accuracy of the sequential and ensemble models for circuits that employ angle and amplitude encoding. However, the performance of these encodings in the parallel model appears to be quite similar, with angle encoding showing a slight improvement in RMSE compared to amplitude encoding, at least for the specific experiments conducted.

When analyzing the results across all architectures, amplitude encoding appears to provide more consistent accuracy compared to angle encoding. This stability suggests that amplitude encoding can reliably represent quantum states from classical data while requiring less complex circuits, making it a robust baseline choice for QNNs.

On the other hand, angle encoding appears to perform better when used with models that have a larger number of parameters. This is suggested by its improved accuracy when combined with the parallel architecture, which contains significantly more parameters than the sequential and ensemble approaches. However, it is important to note that increasing the number of parameters does not necessarily lead to better performance. This is evident in the ensemble model using ansatz 2 and angle encoding, which has a parameter count comparable to the parallel model but yields inferior results. This observation suggests a possible hypothesis: the effectiveness of angle encoding may depend on architectures with higher model capacity. Further studies are needed to investigate this relationship and determine whether more expressive architectures can better leverage angle-based encodings.

Regarding the impact of different ansätze, no significant differences in model accuracy were observed, with only marginal variations when paired with amplitude encoding. However, a more noticeable difference can be seen for ansätze using angle encoding in the sequential and ensemble models, where ansatz 1 appears to yield better results compared to the other circuits. This finding aligns with related studies in which this particular ansatz is considered one of the most recommended due to its lower complexity and ability to express nonlinear functions \cite{b15}.

After analyzing the results, the variation with the lowest RMSE on the PDBind dataset was selected for each quantum architecture, as quantum models demonstrated superior performance on this specific dataset. Table \ref{tbl:qnn-baseline-comparison} presents the performance metrics regarding the accuracy of the selected quantum models compared to the baseline neural network (NN). It can be observed that, for the training dataset and nanobody data, the baseline NN exhibited better accuracy. However, for the PDBind dataset, all quantum models showed an improvement of approximately 20\% over the baseline neural network’s performance.

\begin{table*}[!ht]
\caption{Accuracy comparison between quantum models and baseline neural network}
\label{tbl:qnn-baseline-comparison}
\centering
\scriptsize
\begin{tabular}[t]{lcccc}
\hline
Model&Train - RMSE&Test - RMSE&Nbs - RMSE&PDBind - RMSE\\
\hline
baseline NN & 2.43 & 2.14 & 1.66 & 2.45\\
Sequential (ansatz 3 w/ amplitude) & 2.92 & 2.64 & 2.14 & 1.92\\
Parallel (ansatz 2 w/ angle) & 2.91 & 2.64 & 2.29 & 1.92\\
Ensemble (ansatz 4 w/ amplitude) & 2.92 & 2.65 & 2.37 & 1.94\\
\hline
\end{tabular}
\end{table*}

It is worth noting that the chosen encodings impose inherent expressivity trade-offs. The limited depth and small number of re-upload blocks used in the angle encoding circuits may restrict their ability to represent highly non-linear relationships, while amplitude encoding, although qubit-efficient, essentially acts as a linear model when combined with shallow PQCs. Nonetheless, the fact that QNNs achieved superior performance on the PDBind dataset despite these limitations suggests that even standard, low-expressivity encodings can capture meaningful interaction patterns in realistic protein-ligand data.

Table \ref{tbl:qnn-baseline-comparison-time} presents the performance metrics regarding the training time of the models, expressed in seconds. There is a clear difference in execution time between the baseline NN and all quantum models, with the most striking difference observed in the sequential model, which exhibited an approximate 98\% reduction in execution time compared to the baseline NN.

\begin{table}[ht]
\caption{Comparison of execution time between quantum models and baseline NN}
\label{tbl:qnn-baseline-comparison-time}
\centering
\scriptsize
\begin{tabular}[t]{lcc}
\hline
Model&Execution time (secs.)&Parameters\\
\hline
Baseline NN & 527.95 & 3230\\
Sequential (ansatz 3 w/ amplitude) & 9.81 & 26\\
Parallel (ansatz 2 w/ angle) & 267.11 & 665\\
Ensemble (ansatz 4 w/ amplitude) & 132.61 & 333\\
\hline
\end{tabular}
\end{table}

This result is likely due to the significantly lower number of parameters in all quantum models, which are smaller than those of the baseline NN. On the other hand, this smaller number of degrees of freedom may limit the ability to generalize to new data. \cite{b30}.

However, despite having fewer degrees of freedom, the quantum models demonstrated satisfactory performance in terms of accuracy, achieving superior results on PDBind datasets. These findings pave the way for future studies in which the number of parameters could be increased to assess whether this modification further enhances accuracy while maintaining a reduced training time.

It is important to note that the execution times reported here reflect idealized simulations under analytic conditions and, therefore, should not be interpreted as directly transferable to NISQ devices. Real quantum hardware would introduce sampling noise, finite-shot effects, and gate errors, which could substantially increase training time and affect convergence. Nonetheless, isolating hardware noise at this stage provides a clearer view of the intrinsic learning capacity of QNNs.

\section{Conclusion}
\label{sec:conclusion}

In this study, different quantum neural network architectures based on multilayer perceptrons were proposed, each incorporating distinct ansätze and feature mappings to analyze their impact on model performance. The angle and amplitude feature mappings used were presented, along with the different ansätze applied. The data re-upload technique, employed for angle mapping, was also introduced, and its effect on the depth of these parameterized quantum circuits was discussed. Values related to circuit complexity, such as the number of parameters and the total number of quantum gates, were also reported.

The performance results of each architecture were presented, along with the total number of trainable parameters for each of their variations. For a direct comparison between the quantum models and the baseline NN, the RMSE 
values, execution time, and number of trainable parameters were provided for the quantum models that demonstrated the best performance and the baseline model.

The impact of different ansätze and feature encodings on the proposed architectures was analyzed. A significant difference was observed between the performance of models employing angle-based and amplitude-based quantum
circuits, with amplitude encoding generally achieving better accuracy, except in the parallel architecture. It was hypothesized that the higher number of trainable parameters in the parallel model may have contributed to the
improved performance of angle-mapped circuits, a factor that could be further explored in future studies. On the other hand, no significant differences were observed among the various ansätze that utilized amplitude encoding. Based on these results, ansatz 1 offers the best performance despite being a relatively less complex circuit.

Based on the obtained results, the baseline NN demonstrated higher accuracy in both the training datasets and the nanobodies dataset. However, the quantum models achieved better accuracy on the PDBind dataset, showing an
improvement of approximately 20\% in the RMSE metric.

The superior performance of quantum models on the PDBind dataset, in contrast to their underperformance on the training and nanobody datasets, can likely be attributed to dataset-specific characteristics. PDBind exhibits greater structural diversity and a wider range of binding affinities, offering richer, more complex patterns for the quantum models to learn. This complexity enables QNNs to more effectively leverage quantum phenomena, such as entanglement and interference, to model nonlinear relationships. In contrast, the more homogeneous training and nanobody datasets may lack the variability necessary for quantum models to outperform classical neural networks.

From a biological standpoint, the PDBind dataset is particularly relevant, as it comprises protein–ligand complexes with structural heterogeneity that better reflects the challenges of real-world drug discovery. The lower RMSE achieved by QNNs on this dataset suggests an enhanced ability to capture subtle molecular interaction patterns, which are critical for accurate binding affinity prediction. These results highlight the potential of quantum models to complement classical approaches in tasks such as virtual screening and scoring, especially as quantum hardware continues to evolve.

Regarding execution time, the results indicate that all proposed quantum models ran significantly faster than the baseline NN. This improvement is likely due to the reduced number of parameters in quantum neural networks, which are orders of magnitude smaller than those of the baseline model.

While this work does not introduce new learning paradigms, it demonstrates that QNN can already achieve accuracy comparable to state-of-the-art classical models in predicting protein binding affinity. The improvement observed in the PDBind dataset is particularly significant because it reflects real-world protein-ligand diversity. These findings suggest that bioinformatics applications, particularly binding affinity prediction, may be among the first areas where quantum computing yields meaningful results, paving the way for future integration into pharmaceutical pipelines as quantum hardware advances.

Future work should aim to enhance the generalizability and practical applicability of QNNs in protein binding tasks. This includes increasing model capacity, incorporating transfer learning, and, importantly, exploring more expressive and hardware-efficient encoding schemes, such as adaptive data re-uploading and domain-inspired feature maps, to better harness the representational power of QNNs. Additionally, extending experiments to finite-shot and noisy simulations is essential for realistically estimating performance on NISQ hardware. To this end, strategies such as error mitigation and hardware-efficient circuit optimization should be investigated to address sampling noise and improve robustness under real-world quantum conditions. These combined efforts could help extend the observed advantages of QNNs to a broader range of molecular datasets and protein families.

Additionally, it is worth noting that the dimensionality reduction was inherited from the classical baseline pipeline to ensure comparability and may also contribute to the observed performance. Future work will include ablation studies using the original 41 features to disentangle better the impact of classical preprocessing from the intrinsic performance of quantum models.

\bibliographystyle{unsrtnat}
\bibliography{references}

\end{document}